\documentclass[twocolumn,tighten]{aastex63}

\usepackage{natbib}
\usepackage{multirow}
\usepackage{amsmath}
\usepackage{hyperref}
\hypersetup{citecolor=blue,urlcolor=blue}
\received{Oct 23, 2020}
\accepted{Dec 30, 2020}

\shorttitle{Extended [\ion{C}{2}] structure in the $z = 6.72$ red quasar J1205$-$0000}
\shortauthors{T. Izumi et al.}
\begin{document}
\title{Subaru High-z Exploration of Low-Luminosity Quasars (SHELLQs) XII.\\
Extended [\ion{C}{2}] Structure (Merger or Outflow) in a $z = 6.72$ Red Quasar}

\correspondingauthor{Takuma Izumi}
\email{takuma.izumi@nao.ac.jp}

\author[0000-0002-0786-7307]{Takuma Izumi}
\altaffiliation{NAOJ Fellow}
\affil{National Astronomical Observatory of Japan, 2-21-1 Osawa, Mitaka, Tokyo 181-8588, Japan}
\affil{Department of Astronomical Science, The Graduate University for Advanced Studies, SOKENDAI, 2-21-1 Osawa, Mitaka, Tokyo 181-8588, Japan}
\author[0000-0003-2984-6803]{Masafusa Onoue} 
\affil{Max-Planck-Institut f\"{u}r Astronomie, K\"{o}nigstuhl 17, D-69117 Heidelberg, Germany}
\author{Yoshiki Matsuoka} 
\affil{Research Center for Space and Cosmic Evolution, Ehime University, 2-5 Bunkyo-cho, Matsuyama, Ehime 790-8577, Japan}
\author{Michael A. Strauss}
\affil{Princeton University Observatory, Peyton Hall, Princeton, NJ 08544, USA}
\author[0000-0001-7201-5066]{Seiji Fujimoto}
\affil{Cosmic Dawn Center (DAWN), Copenhagen, Denmark}
\affil{Niels Bohr Institute, University of Copenhagen, Jagtvej 128, 2200 Copenhagen N}
\author{Hideki Umehata}
\affil{RIKEN Cluster for Pioneering Research, 2-1 Hirosawa, Wako, Saitama 351-0198, Japan}
\affil{Institute of Astronomy, Graduate School of Science, The University of Tokyo, 2-21-1 Osawa, Mitaka, Tokyo 181-0015, Japan}
\author{Masatoshi Imanishi}
\affil{National Astronomical Observatory of Japan, 2-21-1 Osawa, Mitaka, Tokyo 181-8588, Japan}
\affil{Department of Astronomical Science, The Graduate University for Advanced Studies, SOKENDAI, 2-21-1 Osawa, Mitaka, Tokyo 181-8588, Japan}
\author{Taiki Kawamuro}
\affil{National Astronomical Observatory of Japan, 2-21-1 Osawa, Mitaka, Tokyo 181-8588, Japan}
\author{Tohru Nagao}
\affil{Research Center for Space and Cosmic Evolution, Ehime University, 2-5 Bunkyo-cho, Matsuyama, Ehime 790-8577, Japan}
\author{Yoshiki Toba}
\affil{Department of Astronomy, Kyoto University, Kitashirakawa-Oiwake-cho, Sakyo-ku, Kyoto 606-8502, Japan}
\affil{Academia Sinica Institute of Astronomy and Astrophysics, 11F of Astronomy-Mathematics Building, AS/NTU, No.1, Section 4, Roosevelt Road, Taipei 10617, Taiwan}
\affil{Research Center for Space and Cosmic Evolution, Ehime University, 2-5 Bunkyo-cho, Matsuyama, Ehime 790-8577, Japan}
\author{Kotaro Kohno}
\affil{Institute of Astronomy, Graduate School of Science, The University of Tokyo, 2-21-1 Osawa, Mitaka, Tokyo 181-0015, Japan}
\affil{Research Center for the Early Universe, Graduate School of Science, The University of Tokyo, 7-3-1 Hongo, Bunkyo-ku, Tokyo 113-0033, Japan}
\author{Nobunari Kashikawa}
\affil{Department of Astronomy, School of Science, The University of Tokyo, 7-3-1 Hongo, Bunkyo-ku, Tokyo 113-0033, Japan}
\author{Kohei Inayoshi}
\affil{Kavli Institute for Astronomy and Astrophysics, Peking University, Beijing 100871, People’s Republic of China}
\author{Toshihiro Kawaguchi}
\affil{Department of Economics, Management and Information Science, Onomichi City University, 1600-2, Hisayamada, Onomichi, Hiroshima, 722-8506, Japan}
\author{Kazushi Iwasawa}
\affil{ICREA and Institut de Ci\`{e}ncies del Cosmos, Universitat de Barcelona, IEEC-UB, Mart\'{i} i Franqu\`{e}s, 1,E-08028 Barcelona, Spain}
\author{Akio K. Inoue}
\affil{Department of Physics, School of Advanced Science and Engineering, Faculty of Science and Engineering, Waseda University, 3-4-1, Okubo, Shinjuku, Tokyo 169-8555, Japan} 
\affil{Waseda Research Institute for Science and Engineering, Faculty of Science and Engineering, Waseda University, 3-4-1, Okubo, Shinjuku, Tokyo 169-8555, Japan}
\author{Tomotsugu Goto}
\affil{Institute of Astronomy and Department of Physics, National Tsing Hua University, Hsinchu 30013, Taiwan}
\author{Shunsuke Baba}
\affil{National Astronomical Observatory of Japan, 2-21-1 Osawa, Mitaka, Tokyo 181-8588, Japan}
\author{Malte Schram}
\affil{Graduate school of Science and Engineering, Saitama University, 255 Shimo-Okubo, Sakura-ku, Saitama City, Saitama 338-8570, Japan} 
\author{Hyewon Suh}
\affil{Subaru Telescope, National Astronomical Observatory of Japan (NAOJ), 650 North A'ohoku Place, Hilo, HI 96720, USA}
\affil{Gemini Observatory/NSF’s NOIRLab, 670 N. A’ohoku Place, Hilo, HI 96720, USA}
\author{Yuichi Harikane}
\affil{Institute for Cosmic Ray Research, The University of Tokyo, 5-1-5 Kashiwanoha, Kashiwa, Chiba 277-8582, Japan}
\author{Yoshihiro Ueda}
\affil{Department of Astronomy, Kyoto University, Kitashirakawa-Oiwake-cho, Sakyo-ku, Kyoto 606-8502, Japan}
\author{John D. Silverman}
\affil{Kavli Institute for the Physics and Mathematics of the Universe (Kavli-IPMU, WPI), The University of Tokyo Institutes for Advanced Study, The University of Tokyo, 5-1-5 Kashiwanoha, Kashiwa, Chiba 277-8583, Japan}
\affil{Department of Astronomy, School of Science, The University of Tokyo, 7-3-1 Hongo, Bunkyo-ku, Tokyo 113-0033, Japan}
\author{Takuya Hashimoto}
\affil{Tomonaga Center for the History of the Universe (TCHoU), Faculty of Pure and Applied Sciences, University of Tsukuba, Tsukuba, Ibaraki 305-8571, Japan}
\author{Yasuhiro Hashimoto}
\affil{Department of Earth Sciences, National Taiwan Normal University, Taipei 11677, Taiwan}
\author{Soh Ikarashi}
\affil{Centre for Extragalactic Astronomy, Department of Physics, Durham University, South Road, Durham DH1 3LE, UK}
\author{Daisuke Iono}
\affil{National Astronomical Observatory of Japan, 2-21-1 Osawa, Mitaka, Tokyo 181-8588, Japan}
\affil{Department of Astronomical Science, The Graduate University for Advanced Studies, SOKENDAI, 2-21-1 Osawa, Mitaka, Tokyo 181-8588, Japan}
\author{Chien-Hsiu Lee} 
\affil{NSF's National Optical-Infrared Astronomy Research Laboratory, 950 North Cherry Avenue, Tucson, AZ 85719, USA}
\author{Kianhong Lee}
\affil{Institute of Astronomy, Graduate School of Science, The University of Tokyo, 2-21-1 Osawa, Mitaka, Tokyo 181-0015, Japan}
\author{Takeo Minezaki}
\affil{Institute of Astronomy, Graduate School of Science, The University of Tokyo, 2-21-1 Osawa, Mitaka, Tokyo 181-0015, Japan}
\author{Kouichiro Nakanishi}
\affil{National Astronomical Observatory of Japan, 2-21-1 Osawa, Mitaka, Tokyo 181-8588, Japan}
\affil{Department of Astronomical Science, The Graduate University for Advanced Studies, SOKENDAI, 2-21-1 Osawa, Mitaka, Tokyo 181-8588, Japan}
\author{Suzuka Nakano}
\affil{Department of Astronomical Science, The Graduate University for Advanced Studies, SOKENDAI, 2-21-1 Osawa, Mitaka, Tokyo 181-8588, Japan}
\author{Yoichi Tamura}
\affil{Division of Particle and Astrophysical Science, Graduate School of Science, Nagoya University, Furo-cho, Chikusa-ku, Nagoya, Aichi 464-8602, Japan}
\author[0000-0002-1860-0886]{Ji-Jia Tang}
\affil{Research School of Astronomy and Astrophysics, Australian National University, Cotter Road, Weston Creek, ACT 2611, Australia}

\begin{abstract}
We present ALMA [\ion{C}{2}] 158 $\mu$m line and far-infrared (FIR) continuum emission observations 
toward HSC J120505.09$-$000027.9 (J1205$-$0000) at $z = 6.72$ with the beam size of $\sim 0\arcsec.8 \times 0\arcsec.5$ (or 4.1 kpc $\times$ 2.6 kpc), 
the most distant red quasar known to date. 
Red quasars are modestly reddened by dust, and are thought to be in rapid transition from an obscured starburst 
to an unobscured normal quasar, driven by  powerful active galactic nucleus (AGN) feedback which blows out a cocoon of interstellar medium (ISM).  
The FIR continuum of J1205$-$0000 is bright, with an estimated luminosity of $L_{\rm FIR} \sim 3 \times 10^{12}~L_\odot$. 
The [\ion{C}{2}] line emission is extended on scales of $r \sim 5$ kpc, greater than the FIR continuum. 
The line profiles at the extended regions are complex and broad (FWHM $\sim 630-780$ km s$^{-1}$). 
Although it is not practical to identify the nature of this extended structure, 
possible explanations include (i) companion/merging galaxies and (ii) massive AGN-driven outflows. 
For the case of (i), the companions are modestly star-forming ($\sim 10~M_\odot$ yr$^{-1}$), 
but are not detected by our Subaru optical observations ($y_{\rm AB,5\sigma} = 24.4$ mag). 
For the case of (ii), our lower-limit to the cold neutral outflow rate is $\sim 100~M_\odot$ yr$^{-1}$. 
The outflow kinetic energy and momentum are both much smaller than 
what predicted in energy-conserving wind models, suggesting that the AGN feedback in this quasar 
is not capable of completely suppressing its star formation. 
\end{abstract}

\section{Introduction}\label{sec1}
It has long been considered that mergers or interactions 
of gas-rich galaxies are the trigger for intense starburst and black hole growth \citep[e.g.,][]{1988ApJ...325...74S,2006ApJS..163....1H}. 
In this scenario, a substantial fraction of the black hole growth takes place inside a dusty starburst, 
during which time the system appears as an obscured active galactic nucleus \citep[obscured AGN,][]{2018ARA&A..56..625H}. 
Subsequent strong AGN feedback, particularly in the form of massive outflows 
\citep[e.g.,][]{2012MNRAS.425L..66M,2014A&A...562A..21C,2016A&A...591A..28C,2019A&A...630A..59B,2019MNRAS.483.4586F,2020A&A...633A.134L}, 
is expected to clear gas and dust from the galaxy, halting further star formation and black hole growth.  
This feedback is also invoked in galaxy evolution models to shape the galaxy mass function 
and ultimately black hole-host galaxy co-evolution relations \citep{2013ARA&A..51..511K}. 

High-redshift quasars are a unique laboratory to test or refine 
our understanding of the early supermassive black hole (SMBH) 
and galaxy formation \citep[e.g.,][]{2019arXiv191105791I}. 
To date, ${}>200$ quasars with absolute ultraviolet (UV) magnitude $M_{\rm 1450} \lesssim -22$ 
are known at $z > 5.7$, most of which were discovered by wide-field optical and near-infrared surveys 
\citep[e.g.,][]{2001AJ....122.2833F,2016ApJS..227...11B,2016ApJ...828...26M,2018PASJ...70S..35M,2018ApJS..237....5M}. 
Sub/millimeter observations have revealed that their host galaxies often possess 
copious amounts of dust ($\sim 10^8~M_\odot$) and gas ($\sim 10^{10}~M_\odot$) 
with high star formation rates (SFR) of $\gtrsim 100$--$1000~M_\odot$ yr$^{-1}$ 
\citep[e.g.,][]{2013ApJ...773...44W,2016ApJ...816...37V,2017ApJ...845..154V,2018ApJ...854...97D}, 
while smaller values are found in lower luminosity quasars 
\citep[e.g.,][]{2013ApJ...770...13W,2015ApJ...801..123W,2017ApJ...850..108W,2018PASJ...70...36I,2019PASJ...71..111I}. 

As a galaxy-overdensity increases the likelihood of interactions and mergers of galaxies, 
deep submillimeter searches for companion star-forming galaxies have been performed 
toward $z \gtrsim 6$ quasars mainly by using the Atacama Large Millimeter/submillimeter Array (ALMA). 
Such efforts tried to detect C$^+$ ${}^2P_{3/2} \rightarrow {}^2P_{1/2}$ 157.74 $\micron$ 
(hereafter [\ion{C}{2}] 158 $\micron$) line and/or rest-frame far-infrared (FIR) continuum emission from companions. 
Indeed, some works found companions \citep[e.g.,][]{2017Natur.545..457D,2017ApJ...836....8T,2017ApJ...850..108W,2019ApJ...882...10N}, 
or very close likely-merging galaxies \citep{2019ApJ...880..157D,2019ApJ...874L..30V,2019ApJ...881L..23B}, 
to the optically-luminous ($M_{\rm 1450} \lesssim -26$ mag) quasars, 
which would support the triggering of quasars by gas-rich major mergers. 
Contrary to this, no overdensity of submillimeter continuum emitters was reported 
for the case of optically less-luminous ($M_{\rm 1450} \gtrsim -25$ mag) quasars \citep{2019PASJ...71..111I}, 
implying that rather secular processes \citep[e.g.,][]{2019MNRAS.482.4846S,2020MNRAS.494.2747M} may also be capable of triggering $z \gtrsim 6$ quasars. 

As massive quiescent galaxies already exist at $z \sim 3-4$ \citep[e.g.,][]{2014ApJ...783L..14S,2020ApJ...898..171E}, 
we also anticipate that AGN feedback is already taking place at higher redshift. 
An outstanding case is the $z = 6.42$ quasar J1148$+$5251, in which a massive 
AGN-driven outflow (outflow rate $\dot{M}_{\rm out} > 1400~M_\odot$ yr$^{-1}$) extending to $r > 10$ kpc 
is detected in the [\ion{C}{2}] 158 $\micron$ emission line \citep{2012MNRAS.425L..66M,2015A&A...574A..14C}. 
However, there is no other single $z>6$ quasar in which [\ion{C}{2}] outflow has
been reported, likely due to inadequate sensitivity. 
To overcome the sensitivity issue, \citet{2019A&A...630A..59B} stacked the [\ion{C}{2}] cubes of 
48 quasars at $4.5 < z < 7.1$, and reported a successful detection of a broad [\ion{C}{2}] wing: 
the inferred outflow rate ($\dot{M}_{\rm out} \sim 100~M_\odot$ yr$^{-1}$) is only modest, 
but may be enough to quench star formation in the central regions of these galaxies. 
On the other hand, \citet{2020arXiv201014875N} performed 
another stacking analysis of [\ion{C}{2}] cubes of 27 $z \gtrsim 6$ quasars, 
revealing no clear evidence of [\ion{C}{2}] outflows \citep[see also][]{2018ApJ...854...97D}. 
Hence the situation is quite controversial these days. 
In either case, however, previous $z \gtrsim 6$ studies are biased toward the unobscured (= blue) quasar population. 

In this regard, one intriguing population of quasars, whose host galaxies 
have not been studied at all at $z > 6$, are the so-called {\it red quasars}. 
Unlike the unobscured population, they are reddened ($E(B-V) > 0.1$) by dust 
at rest-UV \citep[e.g.,][]{2003AJ....126.1131R,2004ApJ...607...60G,2018ApJ...861...37G,2015MNRAS.453.3932R}. 
Red quasars typically have high Eddington ratios and are accompanied 
by broad absorption line (BAL) features indicative of nuclear fast outflows \citep[e.g.,][]{2003AJ....126.1131R,2009ApJ...698.1095U}. 
At lower redshifts like $z \lesssim 2.5$, red quasars often host 
galaxy-scale outflows at various gas phases \citep[e.g.,][]{2018A&A...612A..29B,2019MNRAS.488.4126P,2019MNRAS.489..497Z}. 
In addition, high resolution imaging observations revealed that red quasars tend to be hosted by 
major mergers \citep[e.g.,][]{2008ApJ...674...80U,2015ApJ...806..218G}. 
These properties support an evolutionary scenario, in which red quasars are emerging 
from the merger-induced dusty starburst phase likely by blowing out surrounding medium 
with powerful winds \citep[e.g.,][]{2012ApJ...745..178F}. 
However, as no red quasar has been identified at $z > 5$ primarily 
due to their apparent faintness in the rest-UV, it is not clear if this picture also holds in the first generation of quasars. 

\subsection{Our target: J1205$-$0000} 
Very recently, we successfully discovered two red quasars at $z > 5.8$ \citep{2020PASJ..tmp..230K} 
by matching the deep sample of quasars optically-selected from 
the Subaru Hyper Suprime-Cam (HSC) wide-field survey \citep{2018PASJ...70S...4A}
with mid-infrared photometric data obtained by the Wide-field Infrared Survey Explorer (WISE). 
In this work, we present our ALMA observations of [\ion{C}{2}] 158 $\mu$m line and underlying FIR continuum 
emission of one of these two sources, J120505.09$-$000027.9 (hereafter 1205$-$0000; Figure \ref{fig1} top-right) at $z = 6.7$, 
by far the highest redshift red quasar known to date \citep{2020PASJ..tmp..230K}
\footnote{\citet{2017ApJ...849...91M} performed submm observations toward this quasar by using NOEMA interferometer, 
but did not detect [\ion{C}{2}] emission as the line was out of their frequency coverage.}. 
Without correcting for dust extinction, J1205$-$0000 has a rest-frame UV magnitude of $M_{\rm 1450} = -24.4$ mag \citep{2016ApJ...828...26M}. 
But once the extinction ($E(B-V) = 0.12$ mag) is accounted for, the quasar has $M_{\rm 1450} = -26.1$ mag, with an Eddington ratio of 0.22 \citep{2020PASJ..tmp..230K}: 
here the reddening was estimated by fitting a typical quasar template to the broad band (optical to mid-IR) 
spectral energy distribution with the Small Magellanic Cloud (SMC) dust extinction law. 
J1205$-$0000 hosts prominent \ion{N}{5} and \ion{C}{4} BALs as well, 
confirming the existence of nuclear outflows \citep{2019ApJ...880...77O}. 

In the following parts, we describe ALMA observations in \S~2, 
extended nature of [\ion{C}{2}] emission in \S~3, and possible origins of the extended structures in \S~4. 
We provide our conclusions in \S~5. 
This is the twelfth in a series of publications presenting the results from our high redshift quasar survey, 
the Subaru High-z Exploration of Low-Luminosity Quasars (SHELLQs). 
Throughout this work, we adopt the standard cosmology with $H_0$ = 70 km s$^{-1}$ Mpc$^{-1}$, 
$\Omega_{\rm M} = 0.3$, and $\Omega_{\rm \Lambda} = 0.7$.

\section{ALMA Observations}\label{sec2}
We observed the redshifted [\ion{C}{2}] line and FIR continuum emission of J1205$-$0000 in Band 6 
on 2020 February 26 (ID = 2019.1.00074.S, PI: T. Izumi), with 41 antennas. 
Our observations were conducted in a single pointing 
with a $24\arcsec$ diameter field of view. 
The baseline length ranged from 15.1 m to 783.5 m, 
resulting in a maximum recoverable scale of $\sim 6\arcsec$. 
J1256$-$0547 was observed as a flux and bandpass calibrator 
and J1217$-$0029 was monitored to calibrate the complex gain. 
The total on-source time was 71 minutes. 

The data were processed using \verb|CASA| version 5.6 (\url{https://casa.nrao.edu}). 
All images were reconstructed with the \verb|tclean| task using natural weighting to maximize the sensitivity. 
We averaged several channels to obtain a velocity resolution of 100 km s$^{-1}$, 
which resulted in a 1$\sigma$ channel sensitivity of 0.13 mJy beam$^{-1}$ (beam size = $0\arcsec.77 \times 0\arcsec.48$, P.A. = $-58\arcdeg$). 
Note that we first cleaned the line cube including the continuum emission down to 3$\sigma$ level to determine the line position. 
With this knowledge, we identified the channels free of line emission. 
All these line-free channels were integrated to generate a continuum map (1$\sigma$ = 17.6 $\mu$Jy beam$^{-1}$), 
which was subtracted in the $uv$ plane by using the first-order polynomial before making the final line cube. 

Throughout the paper, only statistical errors are displayed unless mentioned otherwise 
(absolute flux uncertainty $\sim 10\%$; ALMA Cycle 7 Proposer's Guide). 
As the [\ion{C}{2}] emission turned out to fall at around the edge of one of the spectral windows, 
we display spectral data from two contiguous windows when necessary. 
We also used the \verb|MIRIAD| software \citep{1995ASPC...77..433S} for some of the analyses in this paper.

\section{Result: Extended [\ion{C}{2}] emission}\label{sec3}
\begin{figure*}
\begin{center}
\includegraphics[width=\linewidth]{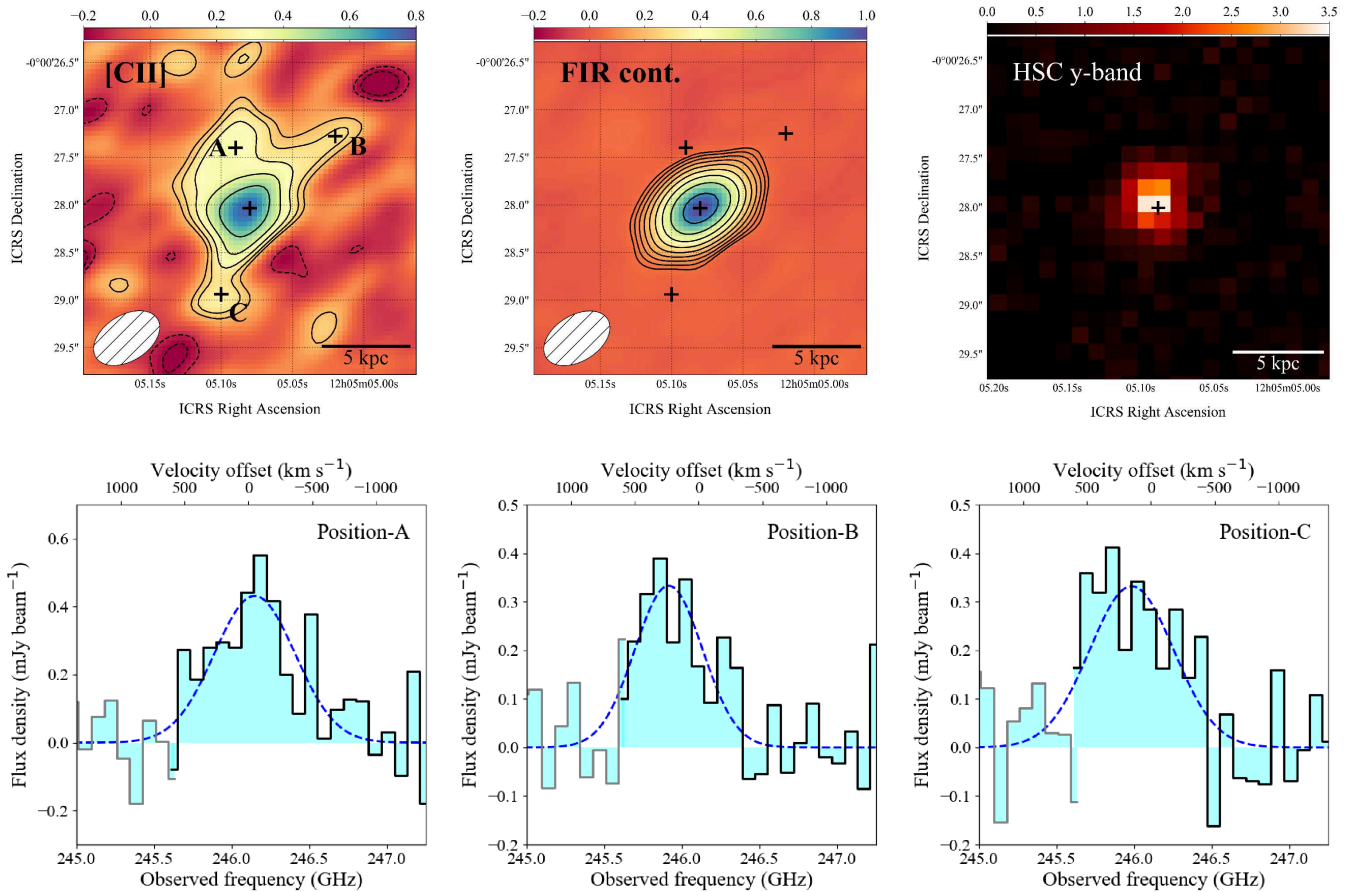}
\caption{
(Top-left) Integrated [\ion{C}{2}] 158 $\micron$ intensity map of J1205$-$0000 (Jy beam$^{-1}$ km s$^{-1}$ unit). 
We integrated over a velocity range of $\pm500$ km s$^{-1}$ relative to the systemic redshift ($z_{\rm [CII]} = 6.7224$), 
which was determined from a double Gaussian fit to the area-integrated spectrum (Figure \ref{fig3}b). 
Contours start at $\pm3\sigma$ (1$\sigma$ = 0.045 Jy beam$^{-1}$ km s$^{-1}$) and increase in powers of $\sqrt{2}$. 
The synthesized beam is shown in the corner. 
The central black plus denotes the continuum peak position. 
Three representative positions to extract off-center spectrum are marked as A, B, and C. 
(Top-middle) 
Spatial distribution of the rest-FIR continuum emission (mJy beam$^{-1}$ unit). 
Contours again start at $\pm3\sigma$ (1$\sigma$ = 17.6 $\mu$Jy beam$^{-1}$) and increase in powers of $\sqrt{2}$. 
The same off-center positions (A--C) are marked. 
(Top-right) Subaru HSC $y$-band map of the same region (arbitrary unit), 
shown as a reference of the optical light distribution of J1205$-$0000. 
(Bottom panels) The [\ion{C}{2}] 158 $\mu$m spectra extracted at the three different off-center positions 
(A--C) with the synthesized beam. 
The velocities are relative to the systemic. 
Results of the single Gaussian fit to these spectra are shown in Table \ref{tbl1}.
}
\label{fig1}
\end{center}
\end{figure*}

The top panels of Figure \ref{fig1} show the distributions of 
the velocity-integrated [\ion{C}{2}] 158 $\micron$ line emission and FIR continuum emission. 
The optical $y$-band image of J1205$-$0000 taken with the Subaru HSC is also shown for a reference. 
Figure \ref{fig2} shows the [\ion{C}{2}] channel maps revealing the global velocity structure. 
We describe details of this emission in the following. 
The relevant physical properties are summarized in Table \ref{tbl1}. 

\subsection{FIR continuum properties}\label{sec3.1}
\begin{deluxetable*}{c|ccc}
\tabletypesize{\small}
\tablecaption{Observed and Derived Properties of J1205$-$0000 \label{tbl1}}
\tablewidth{0pt}
\tablehead{
 & \multicolumn{3}{c}{Area-integrated [\ion{C}{2}] 158 $\mu$m Line Emission} \\ \cline{2-4}
 & \colhead{\multirow{2}{*}{Single Gaussian}} & \multicolumn{2}{c}{Double Gaussian} \\ \cline{3-4}
 & & Core & Wing 
}
\startdata
$z_{\rm [CII]}$ & 6.7230 $\pm$ 0.0003 & 6.7224 $\pm$ 0.0003 & 6.7336 $\pm$ 0.0009 \\ 
FWHM$_{\rm [CII]}$ (km s$^{-1}$) & 536 $\pm$ 26 & 442 $\pm$ 33 & 232 $\pm$ 72 \\ 
$S_{\rm [CII]}\Delta V$ (Jy km s$^{-1}$) & 1.69 $\pm$ 0.07 & 1.49 $\pm$ 0.10 & 0.21 $\pm$ 0.07 \\ 
$L_{\rm [CII]}$ (10$^8$ $L_\odot$) & 18.7 $\pm$ 0.8 & 16.5 $\pm$ 1.0 & 2.4 $\pm$ 0.8 \\ 
SFR$_{\rm [CII]}$ ($M_\odot$ yr$^{-1}$) & 122 $\pm$ 5 & 108 $\pm$ 7 & 16 $\pm$ 6 \\
Size (beam-convolved) & \multicolumn{3}{c}{(1\arcsec.55 $\pm$ 0\arcsec.30) $\times$ (1\arcsec.03 $\pm$ 0\arcsec.17) $=$ (8.3 $\pm$ 1.6) kpc $\times$ (5.5 $\pm$ 0.9) kpc} \\ 
Size (beam-deconvolved) & \multicolumn{3}{c}{(1\arcsec.43 $\pm$ 0\arcsec.36) $\times$ (0\arcsec.77 $\pm$ 0\arcsec.33) $=$ (7.7 $\pm$ 1.9) kpc $\times$ (4.1 $\pm$ 1.8) kpc} \\ \cline{1-4}
 & \multicolumn{3}{c}{[\ion{C}{2}] 158 $\mu$m Line Emission at Selected Positions (Figure \ref{fig1})} \\ 
  & A & B & C  \\ \cline{1-4}
Line center (km s$^{-1}$) & $-$49 $\pm$ 54 & 235 $\pm$ 61 & 151 $\pm$ 69 \\ 
FWHM$_{\rm [CII]}$ (km s$^{-1}$) & 746 $\pm$ 124 & 625 $\pm$ 145 & 775 $\pm$ 162 \\ 
$S_{\rm [CII]}\Delta V$ (Jy beam$^{-1}$ km s$^{-1}$) & 0.34 $\pm$ 0.05 & 0.22 $\pm$ 0.04 & 0.27 $\pm$ 0.05 \\ 
$L_{\rm [CII]}$ (10$^8$ $L_\odot$) & 3.8 $\pm$ 0.6 & 2.5 $\pm$ 0.5 & 3.0 $\pm$ 0.6 \\ \cline{1-4}
 & \multicolumn{3}{c}{Continuum Emission ($T_d$ = 47 K, $\beta$ = 1.6, $\kappa_{\rm 250GHz}$ = 0.4 cm$^2$ g$^{-1}$)} \\ \cline{1-4}
$f_{\rm 1.2mm}$ (mJy) & \multicolumn{3}{c}{1.17 $\pm$ 0.04} \\ 
$L_{\rm FIR}$ (10$^{12}$ $L_\odot$) &\multicolumn{3}{c}{2.7 $\pm$ 0.1} \\ 
$L_{\rm TIR}$ (10$^{12}$ $L_\odot$) & \multicolumn{3}{c}{3.9 $\pm$ 0.1} \\ 
SFR$_{\rm TIR}$ ($M_\odot$ yr$^{-1}$) & \multicolumn{3}{c}{575 $\pm$ 21} \\ 
$M_{\rm dust}$ (10$^8$ $M_\odot$) & \multicolumn{3}{c}{2.0 $\pm$ 0.1} \\ 
Size (beam-convolved) & \multicolumn{3}{c}{(0\arcsec.79 $\pm$ 0\arcsec.02) $\times$ (0\arcsec.52 $\pm$ 0\arcsec.01) $=$ (4.3 $\pm$ 0.1) kpc $\times$ (2.8 $\pm$ 0.1) kpc} \\ 
Size (beam-deconvolved) & \multicolumn{3}{c}{(0\arcsec.25 $\pm$ 0\arcsec.06) $\times$ (0\arcsec.17 $\pm$ 0\arcsec.09) $=$ (1.3 $\pm$ 0.3) kpc $\times$ (0.9 $\pm$ 0.5) kpc} \\ 
\enddata
\tablecomments{The SFR$_{\rm [CII]}$ for the wing component is appropriate if this reflects a star-forming companion galaxy.}
\end{deluxetable*}

The continuum emission ($\lambda_{\rm obs} = 1.2$ mm) comes 
from the central $\sim 1\arcsec$ (see Table \ref{tbl1} for detailed sizes measured with the CASA task \verb|imfit|). 
The continuum peaks at (RA, Dec)$_{\rm ICRS}$ = (12$^{\rm h}$05$^{\rm m}$05$^{\rm s}$.080, $-$00$\arcdeg$00$\arcmin$28$\arcsec$.04), 
which is consistent with the optical quasar position. 
To include all extended emission, we measure the continuum flux density 
with a 2$\arcsec$ diameter circular aperture placed at this peak position, 
resulting in $f_{\rm 1.2mm} = 1.17 \pm 0.04$ mJy. 

We determine the FIR luminosity ($L_{\rm FIR}$; 42.5--122.5 $\micron$) 
and the total IR luminosity ($L_{\rm TIR}$; 8--1000 $\micron$) 
by assuming an optically-thin modified black body with dust temperature $T_d = 47$ K 
and emissivity index $\beta = 1.6$, values which are characteristic of high-redshift luminous quasars \citep{2006ApJ...642..694B} 
and are commonly adopted in the literature \citep[e.g.,][]{2018PASJ...70...36I}. 
After correcting for the contrast and the additional heating effects 
of the cosmic microwave background radiation \citep{2013ApJ...766...13D}, 
we obtain $L_{\rm FIR} = (2.7 \pm 0.1) \times 10^{12}~L_\odot$ and $L_{\rm TIR} = (3.9 \pm 0.1) \times 10^{12}~L_\odot$, respectively. 
Note that there is actually a wide variety in $T_d$ from source to source \citep{2018ApJ...866..159V}. 
If we adopt $T_d = 35$ K, typical for local star-forming galaxies \citep{2012ApJS..203....9U}, 
these values become $\sim 2\times$ smaller. 

With this $L_{\rm TIR}$ ($T_d = 47$ K) and the calibration of \citet{2011ApJ...737...67M}, 
we found a SFR of $575 \pm 21$ $M_\odot$ yr$^{-1}$. 
However, this should be regarded as an upper limit, as we have assumed that 
the FIR emission is due entirely to star formation, even though there may be 
a significant contribution from the AGN itself \citep[e.g.,][]{2017MNRAS.465.1401S}. 
Similarly, by adopting a rest-frame mass absorption coefficient of
$\kappa_\nu = 0.4~(\nu/250{\rm GHz})^\beta$ cm$^2$ g$^{-1}$ \citep{2004A&A...425..109A}, 
we find a dust mass of $M_{\rm dust} = (2.0 \pm 0.1) \times 10^8~M_\odot$.

\subsection{[\ion{C}{2}] line emission}\label{sec3.2}
Figure \ref{fig1} demonstrates that the [\ion{C}{2}] emission is spatially 
more extended than the FIR continuum emission, extending to $r \sim 1\arcsec$ (5.4 kpc)
\footnote{Although the distribution is complex, 
we fit a 2-dimensional elliptical Gaussian to it (CASA task imfit; Table \ref{tbl1}). 
The estimated beam-deconvolved size is $\sim 7.7 \times 4.1$ kpc$^2$, 
which is much larger than that of the FIR continuum.}, and has a complex morphology. 
The extended structure that is directly connected to the continuum-bright core region is real, 
as we see positive signals from several contiguous velocity channels 
measured there (Figure \ref{fig1}; positions A--C), 
over [$-$500, $+$500] km s$^{-1}$ relative to the systemic velocity (defined in the following). 

\begin{figure}
\begin{center}
\includegraphics[width=\linewidth]{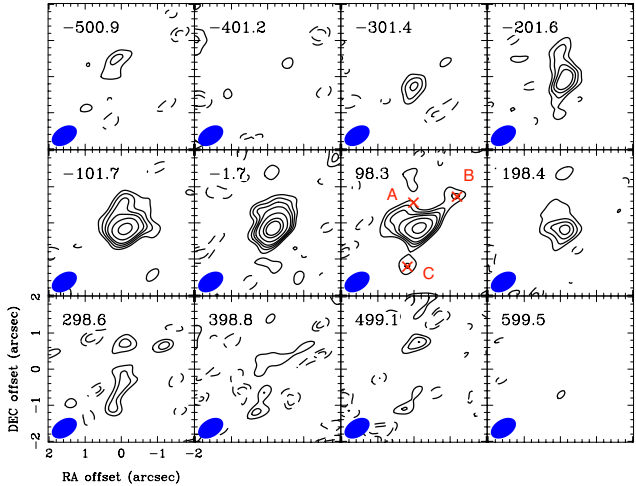}
\caption{
Velocity channel maps of [\ion{C}{2}] line emission of J1205$-$0000. 
Each channel is labelled with its central velocity in km s$^{-1}$. 
The plus signs indicate the FIR continuum peak position. 
The three representative off-center positions (Figure \ref{fig1}) are also marked at the 98 km s$^{-1}$ channel. 
The synthesized beam is plotted in the bottom-left corner. 
Contours are drawn at $-$3, $-$2 (dashed line), 2, 3, 4, 5, 7, 10, 12, and 15$\sigma$ levels (solid line; 1$\sigma$ = 0.13 mJy beam$^{-1}$). 
Extended emission exists at several off-center positions over several contiguous channels. 
}
\label{fig2}
\end{center}
\end{figure}

\begin{figure}
\begin{center}
\includegraphics[width=\linewidth]{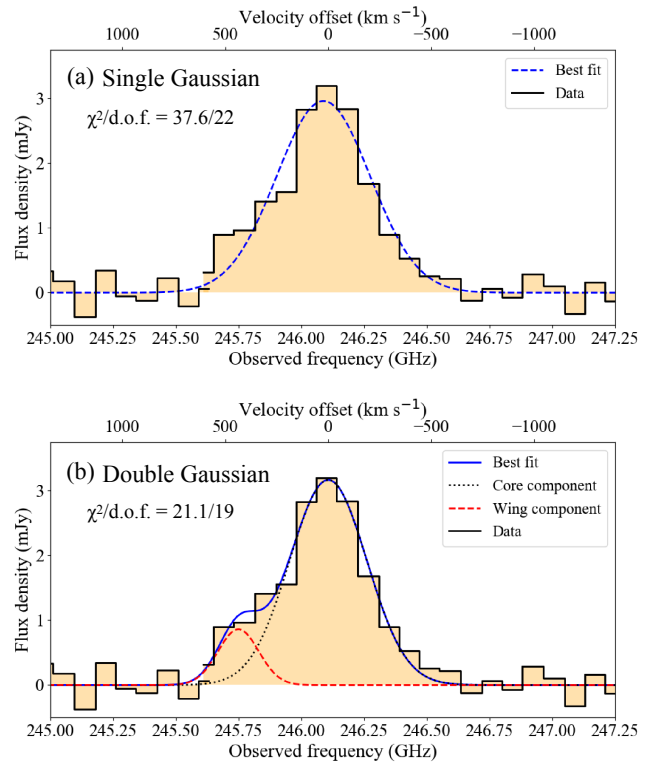}
\caption{
Area-integrated [\ion{C}{2}] 158 $\mu$m spectrum of J1205$-$0000 measured 
over a region of [\ion{C}{2}] integrated-intensity $> 3\sigma$ $\cap$ $r < 1\arcsec$.5. 
The channel-based sensitivity is 0.21 mJy. 
(a) Single Gaussian fit and (b) double Gaussian fit to the observed spectrum.
}
\label{fig3}
\end{center}
\end{figure}

Figure \ref{fig3} shows an area-integrated spectrum measured over the region 
within $1\arcsec.5$ radius of the nucleus in which [\ion{C}{2}] integrated-intensity is detected at more than $3\sigma$. 
We first fit the spectrum with a single Gaussian function (Figure \ref{fig3}a), 
but the observed profile clearly deviates from this model; 
we found $\chi^2$/d.o.f. = 37.6/22 (estimated over $\pm$1200 km s$^{-1}$ range) for this single Gaussian fit. 
A notable deviation can be found at a positive velocity channels of $\sim 300-500$ km s$^{-1}$. 
As the spatially extended regions (Figure \ref{fig2}) have such high positive velocities, 
they must contribute to this {\it excess wing} of the area-integrated spectrum. 
We then performed a {\it double} Gaussian fit (Figure \ref{fig3}b) to account for that excess. 
This works fairly well with a returned $\chi^2$/d.o.f. = 21.1/19. 
With this difference in $\chi^2$, this model is preferred at $>99.9\%$ confidence. 
Note that we also see extended [\ion{C}{2}] emission in negative velocity channels (Figure \ref{fig2}), 
but deeper observations are required to confirm its existence (or another wing at the bluer side of the area-integrated spectrum).

We naively attribute the core component in the double Gaussian fit to the host galaxy itself, 
hence providing its systemic redshift as $z_{\rm [CII]} = 6.7224 \pm 0.0003$. 
With respect to this $z_{\rm [CII]}$, \ion{Mg}{2} emission ($z_{\rm MgII} = 6.699^{0.007}_{-0.001}$) \citep{2019ApJ...880...77O} 
is blueshifted by $-1040^{+310}_{-50}$ km s$^{-1}$. 
In addition, an absorption feature of a higher ionization line \ion{C}{4} of J1205$-$0000 
is further blueshifted by $\sim 2900-7400$ km s$^{-1}$ relative to \ion{Mg}{2} \citep{2019ApJ...880...77O}, 
hence \ion{C}{4} is blueshifted up to $\sim 8500$ km s$^{-1}$ relative to $z_{\rm [CII]}$. 
These clearly manifest the existence of very fast nuclear outflows in this red quasar. 

With the [\ion{C}{2}] core flux and the 1.2 mm continuum flux density (Table \ref{tbl1}), 
we also measure a rest-frame [\ion{C}{2}] equivalent width as $\rm{EW}_{\rm [CII]} = 0.67 \pm 0.05$ $\micron$. 
This is roughly half of the mean $\rm{EW}_{\rm [CII]}$ of local star-forming galaxies \citep[$1.27 \pm 0.53$ $\micron$, e.g.,][]{2013ApJ...774...68D}. 
In optically low-luminosity $z > 6$ quasars (e.g., HSC quasars with $M_{\rm 1450} > -25$ mag) $\rm{EW}_{\rm [CII]}$ is usually $>1$ $\micron$, 
but optically luminous quasars tend to show $\rm{EW}_{\rm [CII]} < 1$ $\micron$ \citep{2019PASJ...71..111I}, 
the latter of which is consistent with the intrinsically luminous nature of J1205$-$0000 
\citep[$M_{\rm 1450} = -26.1$ mag after extinction correction,][]{2020PASJ..tmp..230K}. 

Lastly, we estimate SFR from the [\ion{C}{2}] luminosity of this component (Table \ref{tbl1}), 
giving $108 \pm 7~M_\odot$ yr$^{-1}$ with the \citet{2011MNRAS.416.2712D} calibration, which has $\sim 0.3$ dex dispersion. 
Again, this would be an upper limit, as some of the [\ion{C}{2}] emission may be due to the quasar itself (i.e., narrow line region). 
Indeed, $L_{\rm [CII]}$ shows a marginal correlation with AGN power \citep{2019PASJ...71..111I}.

\subsection{Comparison to the other $z \gtrsim 6$ quasars}\label{sec3.3}
\begin{figure}
\begin{center}
\includegraphics[width=\linewidth]{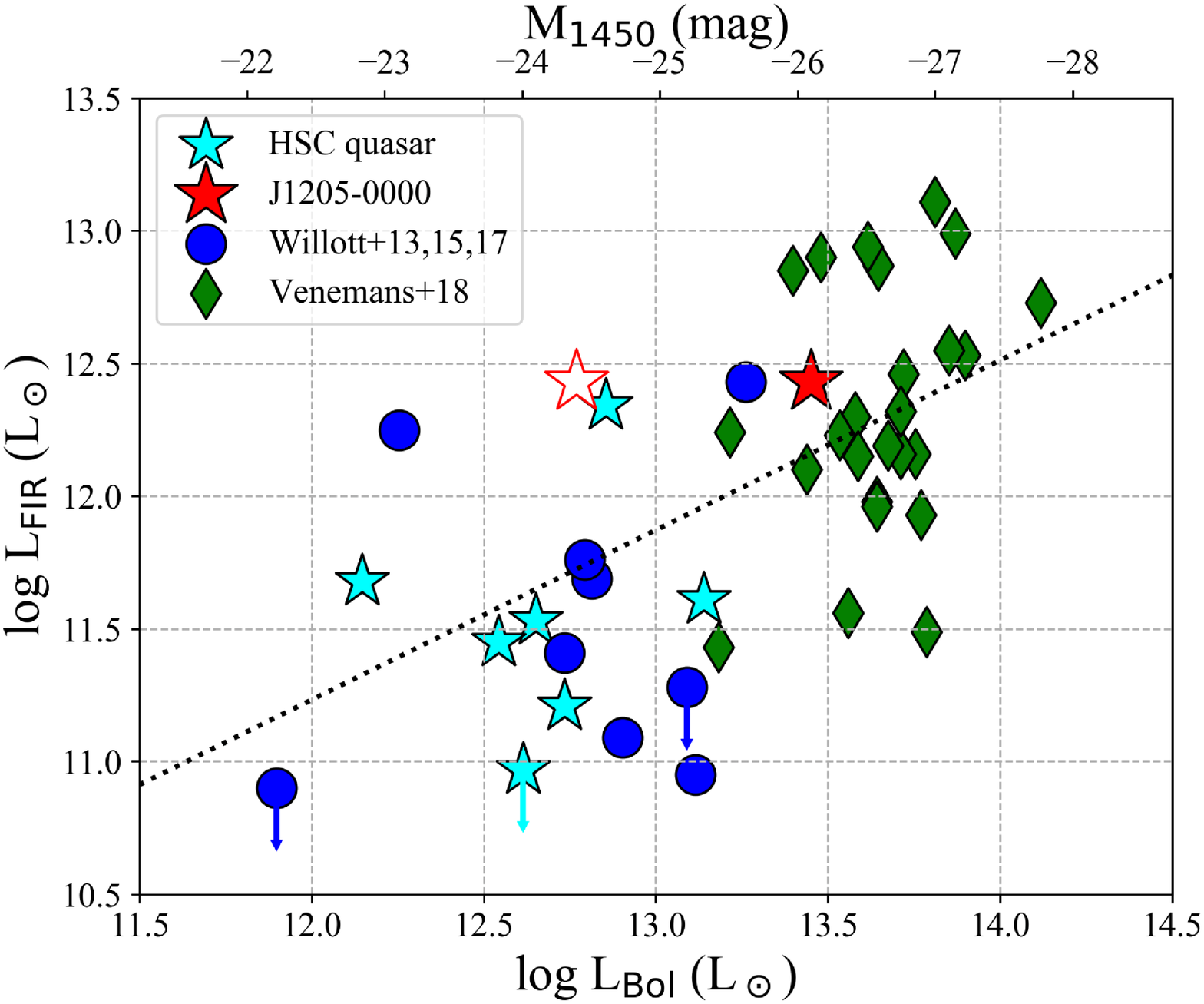}
\caption{
Quasar bolometric luminosity ($L_{\rm Bol}$) vs FIR continuum luminosity ($L_{\rm FIR}$) relationship for $z \gtrsim 6$ quasars. 
The corresponding $M_{\rm 1450}$ is also shown. 
Literature data of some optically luminous quasars \citep{2018ApJ...866..159V}, as well as low-luminosity quasars 
\citep{2013ApJ...770...13W,2015ApJ...801..123W,2017ApJ...850..108W} including our HSC quasars \citep{2018PASJ...70...36I,2019PASJ...71..111I}, are shown. 
J1205$-$0000 after (before) extinction correction is indicated by the red filled (open) star. 
The diagonal line shows our linear regression fit to these objects excluding those with upper limits on $L_{\rm FIR}$. 
}
\label{fig4}
\end{center}
\end{figure}

We briefly compare the quasar nuclear bolometric luminosity ($L_{\rm Bol}$) and $L_{\rm FIR}$ of a sample of $z \gtrsim 6$ quasars in Figure \ref{fig4}. 
Here we adopted the bolometric correction factor 4.4 from 1450 {\AA} luminosity \citep{2006ApJS..166..470R} to compute $L_{\rm Bol}$. 
The same assumptions presented in \S~\ref{sec3.1} are made to derive $L_{\rm FIR}$. 
Optically luminous quasar data is compiled from a recent survey work of \citet{2018ApJ...866..159V}, 
whereas we collected the data of relatively low-luminosity quasars from \citet{2013ApJ...770...13W,2015ApJ...801..123W,2017ApJ...850..108W} 
and \citet{2018PASJ...70...36I,2019PASJ...71..111I}. 

Although there is admittedly a wide spread in $L_{\rm FIR}$ for a given $L_{\rm Bol}$, 
J1205$-$0000 shows a characteristic $L_{\rm FIR}$ to the optically luminous quasars once its dust extinction is corrected. 
If we assume that a quasar-phase happens during a longer timescale starburst-phase 
\citep[e.g.,][]{2007ApJ...671.1388D,2008ApJS..175..356H,2016A&A...587A..72B}, 
there was once a growing {\it obscured AGN} phase in J1205$-$0000 which was embedded in this starburst host galaxy. 

At the lower quasar luminosity regime, i.e., $L_{\rm Bol} < 10^{13}~L_\odot$, 
there are only two quasars (out of twelve) having comparable $L_{\rm FIR}$ to J1205$-$0000, 
namely J2239$+$0207 \citep{2019PASJ...71..111I} and VIMOS2911 \citep{2017ApJ...850..108W}. 
J2239$+$0207 is likely to have a close [\ion{C}{2}] companion galaxy, 
which may have triggered its intense starburst \citep{2019PASJ...71..111I}. 
VIMOS2911, which was originally discovered by \citet{2015ApJ...798...28K}, seems to be an outlier in this plane 
as no companion was reported \citep{2017ApJ...850..108W}, 
as well as its narrow [\ion{C}{2}] line profile (FWHM = 264 km s$^{-1}$) indicates an absence of a merger-like active event. 
The remaining low-luminosity quasars all show $L_{\rm FIR} < 10^{12}~L_\odot$ (some show as low as $< 10^{11}~L_\odot$), 
which is clearly not the case of the optically luminous quasars. 

Indeed, we found a positive correlation between $\log L_{\rm Bol}$ and $\log L_{\rm FIR}$, 
after excluding the objects with upper limits on $L_{\rm FIR}$, as 
\begin{equation}
\log \left( \frac{L_{\rm FIR}}{L_\odot} \right) = (0.64 \pm 0.03) \times \log \left( \frac{L_{\rm Bol}}{L_\odot} \right) + (3.55 \pm 0.45).
\end{equation}
A Spearman correlation coefficient for this $\log L_{\rm Bol}$--$\log L_{\rm FIR}$ data is 0.56 ($p$-value = $2.0 \times 10^{-4}$). 
Although it is still a marginal correlation, this implies that a black hole and its host galaxy are actually co-evolving at this high redshift. 
In addition, if we assume a common star-formation efficiency over the whole quasars used here, 
this correlation suggests that an optically luminous quasar tends to form in a gas richer galaxy. 
Note that \citet{2018ApJ...866..159V} did not find a significant correlation 
between these two luminosities, likely because they only studied optically luminous quasars.

\section{Nature of the extended [\ion{C}{2}] emission}\label{sec4}
We now turn to the physical origin of the extended [\ion{C}{2}] structure, 
which also contributes to the [\ion{C}{2}] wing in the area-integrated spectrum. 
Major possibilities include {\it companion/merging galaxy} and {\it cold outflow}. 
However, as detailed later, it is not possible to distinguish these two scenarios with the current dataset. 
Thus, we discuss both of these scenarios and provide relevant properties, respectively in the following. 
Note that other key properties including the early co-evolution of this quasar will be discussed 
in our forthcoming paper, along with those of the other HSC quasars observed in our ALMA Cycle 7 program (T. Izumi et al. in preparation).

\subsection{Companion/merging galaxies?}\label{sec4.1}
Major mergers of gas-rich galaxies has been considered to be a promising driver of quasar activity 
\citep[e.g.,][]{1988ApJ...325...74S,2005Natur.433..604D,2006ApJS..163....1H}. 
High resolution and/or high sensitivity observations at the low redshift universe 
indeed show an enhanced AGN fraction in major merger systems \citep[e.g.,][]{2011MNRAS.418.2043E,2011ApJ...743....2S,2018PASJ...70S..37G,2018Natur.563..214K}. 
It is also noteworthy that (intrinsically luminous) red quasars tend to be hosted by major merger systems \citep[e.g.,][]{2008ApJ...674...80U,2015ApJ...806..218G}. 
In this context, it is intriguing that recent sensitive ALMA observations of rest-FIR continuum and/or [\ion{C}{2}] emission 
started to uncover star-forming companion galaxies to some $z \gtrsim 5-6$ luminous quasars \citep[e.g.,][]{2011ApJ...739L..34W,2017ApJ...836....8T,2017Natur.545..457D,2017ApJ...850..108W,2019ApJ...882...10N}, 
with some of these companions very close to, or even merging with, the main quasar \citep{2019ApJ...880..157D,2019ApJ...874L..30V,2019ApJ...881L..23B,2020ApJ...904..130V}. 

We found that the [\ion{C}{2}] spatial distribution of J1205$-$0000 including the channel maps (Figures \ref{fig1} and \ref{fig2}) well resembles 
that of J1342$+$0928 at $z = 7.54$ \citep{2019ApJ...881L..23B}, which is considered to be an ongoing merger system. 
Therefore it is plausible that the extended [\ion{C}{2}] structure of J1205$-$0000 is 
due to very close companion or merging galaxies. 
As the [\ion{C}{2}] emission extends toward multiple directions, 
there could be corresponding multiple mergers in J1205$-$0000. 
This picture fits into an evolutionary scenario in which a red quasar 
corresponds to an emergent phase from a merger-induced dusty starburst. 
A high fraction of merger event in red quasar-host galaxies \citep[e.g.,][]{2008ApJ...674...80U,2015ApJ...806..218G} 
measured at lower redshifts also supports this scenario. 
Although we need higher resolution and sensitivity [\ion{C}{2}] data 
to better understand the dynamical nature of this system, 
J1205$-$0000 has a great potential to be an ideal laboratory to study physical properties 
of merging galaxies that would eventually evolve into an unobscured luminous quasar. 

Note that we do not find any companion in our optical $z$- and $y$-band maps ($\sim 0.\arcsec8$ resolution) with $5\sigma$ limiting magnitudes of 
($z_{\rm AB}$, $y_{\rm AB}$) = (25.1, 24.4), taken by HSC \citep[][see also Figure \ref{fig1}]{2016ApJ...828...26M}. 
Although previous ALMA observations actually uncovered optically-unidentified 
[\ion{C}{2}] or FIR continuum emitters around some $z > 6$ quasars \citep[e.g.,][]{2017Natur.545..457D,2019ApJ...881..163M}, 
the situation appears to be different in J1205$-$0000. 
Those known companions are typically bright at FIR with SFR$_{\rm IR}$ of $> 100~M_\odot$ yr$^{-1}$. 
Hence their stellar emission would suffer severe dust extinction, 
which can make them optically unidentifiable \citep{2019ApJ...881..163M}. 
Contrary to this, the FIR continuum emission of J1205$-$0000 is undetected ($<1.5\sigma$) 
at the regions with the spatially extended [\ion{C}{2}] emission (Figure \ref{fig1}). 
An inferred $3\sigma$ upper limit on SFR$_{\rm IR}$ by assuming a modest 
$T_d$ of 35 K \citep[typical value of local star-forming galaxies,][]{2012ApJS..203....9U} is only $\sim 10~M_\odot$ yr$^{-1}$. 
Hence a corresponding dust extinction is not so significant for the potential companions of J1205$-$0000. 
Rather, an insufficient optical survey depth would simply matter. 
If we suppose Lyman break galaxies (LBGs) are responsible for the potential companions, 
our $y$-band observation ($5\sigma_{\rm AB} = 24.4$ mag) only probes 
the bright-end of a UV luminosity function of LBGs at $z \sim 7$ \citep{2018PASJ...70S..10O}. 
In this regard, much deeper data from future facilities such as James Webb Space Telescope (JWST) is needed 
to conclusively identify the companions.

\subsection{Cold neutral outflows?}\label{sec4.2}
The extended and complex gas morphology of J1205$-$0000 also reminds us that of J1148$+$5251, 
the only other known $z > 6$ quasar with prominent [\ion{C}{2}] outflows. 
While J1148$+$5251 is $\sim 5\times$ brighter than J1205$-$0000 at UV 
and the observed [\ion{C}{2}] extent is as large as $r \sim 30$ kpc \citep{2015A&A...574A..14C}, 
this hints at the existence of cold outflows also in J1205$-$0000. 
In addition, the off-center [\ion{C}{2}] spectra are all broad (FWHM $\gtrsim 600-800$ km s$^{-1}$). 
These are even broader than that of the core component, 
as well as than those of known companion galaxies of $z > 6$ quasars 
\citep[$\lesssim 400-500$ km s$^{-1}$, e.g.,][]{2019ApJ...882...10N}, 
which would prefer the outflow scenario. 
Thus, we discuss potential outflow properties hereafter by assuming 
that the extended structure is formed fully due to cold outflows. 
Given that the (potential) outflows are both spatially and spectrally resolved, we can 
measure the dynamical timescales of each [\ion{C}{2}] blob defined in Figure \ref{fig1} (A, B, C). 
With these timescales, we follow the standard prescription \citep[e.g.,][]{2019A&A...630A..59B} 
to estimate outflow rate, kinetic power, and momentum (Table \ref{tbl2}). 

We define the outflow rate as $\dot{M}_{\rm out} = M_{\rm out} v_{\rm out}/R_{\rm out}$, 
where $M_{\rm out}$ is outflowing mass, $v_{\rm out}$ is outflow velocity, 
and $R_{\rm out}$ is the distance from the continuum peak. 
For $v_{\rm out}$, we adopt $|\Delta v| + {\rm FWHM}/2$, 
which combines the shift of the velocity centroid from the systemic and the FWHM of each blob. 
$R_{\rm out}$ is directly measured from Figure \ref{fig1} as 3.5, 6.4, and 5.1 kpc for the blobs A--C, respectively. 
Note that $v_{\rm out}$ is projected along the line of sight, and $R_{\rm out}$ is projected in the plane of the sky, thus, 
 $\tau_{\rm out}$ should be corrected by a factor $|\tan \alpha|^{-1}$, 
where $\alpha$ is the angle between our line-of-sight and the outflow direction of each blob. 
While we do not know $\alpha$ a priori, the average of this factor over all angles is unity. 

We calculate the outflow mass in neutral hydrogen gas $M_{\rm out}$ as
\begin{equation}
\begin{split}
\frac{M_{\rm out}}{M_\odot} &= 0.77 \left( \frac{0.7L_{\rm [CII]}}{L_\odot} \right) \left( \frac{1.4 \times 10^{-4}}{X_{\rm C^+}} \right) \\
& \times \frac{1 + 2e^{-91/T_{\rm ex}} + n_{\rm crit}/n}{2 e^{-91/T_{\rm ex}}}
\end{split}
\end{equation}
where $X_{\rm C^+}$ is the C$^+$ fraction per H atom, $T_{\rm ex}$ is the excitation temperature in K, 
$n$ is the gas density, and $n_{\rm crit}$ is the line critical density ($\sim 3 \times 10^3$ cm$^{-3}$). 
As the actual gas density is unknown, we here assume the high density limit ($n \gg n_{\rm crit}$) to provide a {\it lower limit} on $M_{\rm out}$. 
The factor 0.7 in the first parenthesis indicates a typical fraction of [\ion{C}{2}] arising from photodissociation regions (PDRs; i.e., the remaining 30\% originates from \ion{H}{2} regions). 
For consistency with previous works \citep[e.g.,][]{2012MNRAS.425L..66M,2015A&A...574A..14C}, 
we adopt $X_{\rm C^+} = 1.4 \times 10^{-4}$ and $T_{\rm ex} = 200$ K, 
which are characteristic to PDRs \citep[e.g.,][]{1997ARA&A..35..179H}, 
but varying $T_{\rm ex} \in [100, 1000]$ K changes the inferred $M_{\rm out}$ by only $\sim 20\%$.  

\begin{deluxetable*}{c|cccccc}
\tabletypesize{\small}
\tablecaption{Possible Galaxy-scale [\ion{C}{2}] Outflow Properties of J1205$-$0000 \label{tbl2}}
\tablewidth{0pt}
\tablehead{
 & $v_{\rm out}$ & $M_{\rm out}$ & $\tau_{\rm out}$ & $\dot{M}_{\rm out}$ & $\dot{E}_{\rm out}$ & $\dot{P}_{\rm out}/\dot{P}_{\rm AGN}$\\ 
Blob ID & (km s$^{-1}$) & (10$^8$ $M_\odot$) & (10$^7$ yr) & ($M_\odot$ yr$^{-1}$) & (10$^8$ $L_\odot$) & (\%)
}
\startdata
A & 422 $\pm$ 82 & 3.8 $\pm$ 0.6 & 0.8 $\pm$ 0.2 & 45 $\pm$ 11 & 6.6 $\pm$ 2.4 & 4.3 $\pm$ 1.4 \\ 
B & 547 $\pm$ 95 & 2.4 $\pm$ 0.5 & 1.1 $\pm$ 0.2 & 21 $\pm$ 5 & 5.1 $\pm$ 1.8 & 2.6 $\pm$ 0.8 \\ 
C & 538 $\pm$ 106 & 2.9 $\pm$ 0.5 & 0.9 $\pm$ 0.2 & 31 $\pm$ 8 & 7.5 $\pm$ 2.9 & 3.9 $\pm$ 1.3 \\ 
\enddata
\end{deluxetable*}

Interestingly, the (projected) flow times of the blobs A--C are all clustered around $\sim 10^7$ yr, 
which suggests that these outflows were launched at roughly the same time. 
Summing over these three blobs, the total atomic outflow mass and outflow rate are 
$M_{\rm out,A-C} = 9.0 \pm 0.9) \times 10^8~M_\odot$ and $\dot{M}_{\rm out,A-C} = 97 \pm 15$ $M_\odot$ yr$^{-1}$, respectively. 
The summed outflow rate is $\sim 8\times$ greater than a (current) BH accretion rate 
of J1205$-$0000 ($\dot{M}_{\rm BH} \sim 12.5~M_\odot$ yr$^{-1}$), 
which is estimated from the quasar bolometric luminosity 
\citep[$L_{\rm AGN} = 8.3 \times 10^{46}$ erg s$^{-1}$,][]{2020PASJ..tmp..230K} 
with a canonical radiative efficiency of $\varepsilon = 0.1$. 

Note that, however, our (possible) outflow rate only refers to the neutral atomic component, 
while a significant fraction of the outflowing gas would be in a molecular phase. 
Indeed,  \citet{2019MNRAS.483.4586F} found in local galaxies that the total mass-flow rate is $\sim 0.5$ dex larger than the atom-only value. 
As our computation provided a lower limit on $M_{\rm out}$, and we measured rates only at the three selected positions, 
J1205$-$0000 is likely to have a (possible) total outflow rate of $\dot{M}_{\rm tot} \gtrsim 300~M_\odot$ yr$^{-1}$. 
This is factor $\gtrsim 3$ greater than the [\ion{C}{2}]-based SFR, and it is not typical for star formation 
to drive a wind at a much greater rate than the SFR \citep[e.g.,][]{2014A&A...562A..21C}. 
We therefore naively suppose that the quasar itself powers this outflow.  
We emphasize that J1205$-$0000 is the highest redshift {\it red quasar} known to date. 
Therefore, it is intriguing that the existence of outflows on the scale of the host galaxy is hinted at here, 
suggesting that the surrounding interstellar medium is actively being blown out, as predicted in SMBH/galaxy evolution scenarios \citep[e.g.,][]{2006ApJS..163....1H}. 

We found that the peak flux density of the red wing is $\sim 30\%$ of that of the core component, 
while \citet{2012MNRAS.425L..66M} found a broad-to-narrow peak 
flux density ratio of $\sim 15\%$ for the case of J1148$+$5251. 
The (possible) outflow velocity is much smaller in J1205$-$0000 than J1148$+$5251, 
but we may be limited by the sensitivity of our measurement to detect faster components. 
Interestingly, a similar line profile to J1205$-$0000 is seen in cold outflows of lower redshift obscured quasars 
\citep[e.g.,][]{2018A&A...612A..29B,2018ApJ...856L...5F}. 
Hence suppose for example that the high-peak wing is due to a dense gas 
that once obscured the nucleus in a shell-like form. 
It had a high wind velocity when it was launched, but is now decelerated after traveling a long way from the center. 
As time progresses and the quasar becomes less reddened, the outflow will continue to slow down. 
It would be hard to identify such a low-velocity component as an outflow unless it is spatially resolved. 
Indeed, spatially extended but low-velocity ($|\Delta v| < 200$ km s$^{-1}$) structures 
are also found in J1148$+$5251 \citep{2015A&A...574A..14C}. 

If the extended structures were formed genuinely due to outflows, 
both the outflow kinetic power 
\begin{equation}
\dot{E}_{\rm out} = \frac{1}{2} \dot{M}_{\rm out} v^2_{\rm out}
\end{equation}
and the momentum load 
\begin{equation}
\dot{P}_{\rm out}/\dot{P}_{\rm AGN} = \frac{\dot{M}_{\rm out} \times v_{\rm out}}{L_{\rm AGN}/c}
\end{equation} 
are significantly smaller (Table \ref{tbl2}) than what energy-conserving AGN feedback models typically predict
\citep[$\dot{E}_{\rm out} \sim 0.05 \times L_{\rm AGN}$, $\dot{P}_{\rm out}/\dot{P}_{\rm AGN} \sim 20$, e.g.,][]{2015ARA&A..53..115K}.  
The small values found here are comparable to those of J1148$+$5251 \citep{2015A&A...574A..14C}, 
implying that these may be typical among both red and blue $z \gtrsim 6$ quasars. 
Hence, the nuclear winds contribute only a small fraction of the galaxy-scale feedback 
if we apply this energy-conserving feedback scenario. 
This indicates that, without additional feedback from star formation, 
the AGN feedback is not able to completely suppress star formation over the entire host galaxy.

\section{Summary}\label{sec5}
We have presented ALMA observations of [\ion{C}{2}] line and underlying FIR continuum 
emission of the highest redshift red quasar known to date, J1205$-$0000 at $z = 6.72$, 
which was discovered in our deep optical imaging survey with the Subaru Hyper Suprime-Cam (HSC). 
Although it appears as a low-luminosity quasar ($M_{\rm 1450} = -24.4$ mag), 
its intrinsic luminosity is as high as $M_{\rm 1450} = -26.1$ mag once its dust extinction is corrected. 
A red quasar population has been considered as an emergent phase from a dusty starburst 
by blowing out its surrounding dense obscuring medium. 
The main findings of this paper are summarized as follows. 

\begin{itemize} 
\item[1.] We successfully detected the [\ion{C}{2}] line and FIR continuum emission. 
The [\ion{C}{2}] emission is spatially more extended (over $r \sim 5$ kpc scale) than the continuum emission. 
The area-integrated [\ion{C}{2}] line profile is complex and is characterized by two components, 
i.e., a main core component that is attributed to the host galaxy itself and an additional red wing component. 
Summing up these two, we found the line luminosity of $L_{\rm [CII]} = (1.9 \pm 0.1) \times 10^9~L_\odot$. 
The FIR continuum emission is also bright as $L_{\rm FIR} = (2.7 \pm 0.1) \times 10^{12}~L_\odot$ ($T_d = 47$ K, $\beta = 1.6$). 
\item[2.] The SFR inferred from the continuum luminosity and [\ion{C}{2}] core luminosity 
is $575 \pm 21~M_\odot$ yr$^{-1}$ and $108 \pm 7~M_\odot$ yr$^{-1}$, respectively. 
Note that, however, these would be regarded as upper limits as we neglect a contamination from this intrinsically luminous quasar itself. 
\item[3.] Based on the [\ion{C}{2}] equivalent width and $L_{\rm Bol}$--$L_{\rm FIR}$ relation, 
we found that J1205$-$0000 shows consistent properties to $z \gtrsim 6$ optically luminous quasars 
once its dust extinction is corrected. 
We also found a marginal positive correlation between $L_{\rm Bol}$ and $L_{\rm FIR}$, 
implying a black hole and its host galaxy are co-evolving. 
\item[4.] Although it is not practical to identify the physical origin of 
the spatially extended [\ion{C}{2}] emission with the current dataset, 
possible explanations include (i) companion/merging galaxies and (ii) cold outflows. 
Both of these appear fascinating as they well fit into the merger-induced galaxy evolution scenario. 
\item[5.] If the extended structure is due to (multiple) companion/merging galaxies, 
this indicates that this red quasar indeed emerges with a merger event. 
While no counterpart was identified in our optical imaging survey with Subaru/HSC 
(e.g., $y$-band 5$\sigma$ limiting magnitude = 24.4 mag), 
dust extinction would not be the prime cause of this 
given the non-detection of FIR continuum emission. 
Rather, we may still not be sensitive enough to detect {\it normal} galaxies like LBGs at that high redshift. 
Deeper imaging observations by future facilities like JWST are necessary to search for the possible companion/merging galaxies. 
\item[6.] If the extended structure is due to cold outflows, we can argue 
that this red quasar is in a key transition phase from a dusty starburst by blowing our its surrounding medium. 
On the other hand, our lower-limit on the cold neutral outflow is only $\sim 100~M_\odot$ yr$^{-1}$, 
and we found that the outflow kinetic energy and momentum are both much smaller than 
what predicted in energy-conserving wind models. 
This suggests that the AGN feedback in this quasar, if genuinely exists, 
is not sufficient to suppress star-formation of the host galaxy. 
\end{itemize}

In this work, we revealed some intriguing host galaxy properties of this red quasar. 
This red quasar was originally identified by matching our low-luminosity quasar sample 
and the mid-IR photometric data obtained by WISE \citep{2020PASJ..tmp..230K}. 
As the HSC survey has still been on-going and our sample of $z = 6-7$ low-luminosity quasars 
has been growing, we would be able to find other red quasars at this high redshift. 
They should be a good sample for extensive follow-up studies by, e.g., ALMA and JWST, 
to better understand the early black hole growth and co-evolution. 

\acknowledgments 
We appreciate the anonymous referee's thorough reading 
and very constructive comments that improved this paper greatly. 
This paper makes use of the following ALMA data: 
ADS/JAO.ALMA\#2019.1.00074.S. 
ALMA is a partnership of ESO (representing its member states), 
NSF (USA) and NINS (Japan), together with NRC (Canada), 
MOST and ASIAA (Taiwan), and KASI (Republic of Korea), 
in cooperation with the Republic of Chile. 
The Joint ALMA Observatory is operated by ESO, AUI/NRAO and NAOJ. 

The Hyper Suprime-Cam (HSC) collaboration includes the astronomical communities 
of Japan and Taiwan, and Princeton University. 
The HSC instrumentation and software were developed by the National Astronomical Observatory of Japan (NAOJ), 
the Kavli Institute for the Physics and Mathematics of the Universe (Kavli IPMU), 
the University of Tokyo, the High Energy Accelerator Research Organization (KEK), 
the Academia Sinica Institute for Astronomy and Astrophysics in Taiwan (ASIAA), and Princeton University. 
Funding was contributed by the FIRST program from the Japanese Cabinet Office, 
the Ministry of Education, Culture, Sports, Science and Technology (MEXT), 
the Japan Society for the Promotion of Science (JSPS), 
the Japan Science and Technology Agency (JST), the Toray Science Foundation, 
NAOJ, Kavli IPMU, KEK, ASIAA, and Princeton University. 

T.I. is supported by the ALMA Japan Research Grant of the NAOJ ALMA Project, NAOJ-ALMA-249. 

\software{CASA \citep{2007ASPC..376..127M}, 
MIRIAD \citep{1995ASPC...77..433S}, 
astropy \citep{astropy:2013,astropy:2018}.}

\bibliography{Izumi2021a_ref}

\end{document}